\documentclass[aps,prc,twocolumn,showpacs,nofootinbib,aps,letterpaper]{revtex4-1}

\usepackage{amsmath}
\usepackage{amsfonts,amssymb}
\usepackage{multirow}
\usepackage[dvips]{graphicx}
\usepackage{float}
\usepackage{appendix}
\usepackage{color}
\usepackage{url}
\usepackage{subfigure}

\def\beq{\begin{equation}}
\def\eeq{\end{equation}}
\def\bea{\begin{eqnarray}}
\def\eea{\end{eqnarray}}

\def\nl{\nonumber\\}

\def\ppbar{\bar p p}
\def\nnbar{\bar n n}
\def\NNbar{\bar N N}
\newcommand{\eebar}{e^+e^-}

\RequirePackage{xspace} \allowdisplaybreaks

\begin{document}

\title{Origin of the structures observed in $e^+e^-$ annihilation into multipion states\\
around the $\bar pp$ threshold} 
\author{Johann Haidenbauer$^1$}
\author{Christoph Hanhart$^1$} 
\author{Xian-Wei Kang$^{1}$}
\author{Ulf-G. Mei{\ss}ner$^{2,1}$}
\affiliation{$^1$Institute for Advanced Simulation, J\"ulich
Center for Hadron Physics, and Institut f\"ur
Kernphysik, Forschungszentrum J\"ulich, D-52425 J\"ulich, Germany\\
$^2$Helmholtz-Institut f\"ur Strahlen- und Kernphysik and Bethe
Center for Theoretical Physics, Universit\"at Bonn, D-53115 Bonn,
Germany}

\begin{abstract}
We analyze the origin of the structures observed in the reactions $e^+e^-\to 3(\pi^+\pi^-)$, $2(\pi^+\pi^-\pi^0)$, 
$\omega\pi^+\pi^-\pi^0$, and $e^+e^-\to 2(\pi^+\pi^-)\pi^0$ 
around the antiproton-proton ($\bar pp$) threshold. We calculate the contribution of the two-step
process $e^+e^-\to \bar NN \to$ multipions to the total reaction amplitude. The amplitude for
$e^+e^-\to \bar NN$ is constrained from near-threshold data on the $e^+e^-\to \bar pp$ 
cross section and the one for $\bar NN \to$ multipions can be likewise fixed from 
available experimental information, for all those $5 \pi$ and $6\pi$ states. The resulting 
amplitude for $e^+e^-\to$ multipions turns out to be large enough to play a role for the considered $e^+e^-$ annihilation
channels and, in three of the four reactions, even allows us to reproduce the data quantitatively near the
$\bar NN$ threshold. 
The structures seen in the experiments emerges then as a threshold effect due to the opening of the
$\bar NN$ channel. 
\end{abstract}

\maketitle

\section{Introduction}
Recently, results from several high-statistics measurements of $\eebar$ annihilation into multipion 
states have been published \cite{FOCUS,BABAR,BABAR2,CMD,CMD2}. One of the striking features in the
data is a pronounced structure in the vicinity of the antinucleon-nucleon ($\NNbar$) threshold 
that appears either as a sharp drop for $e^+e^- \to 3(\pi^+\pi^-)$ \cite{BABAR,CMD} or as a dip
for $e^+e^- \to 2(\pi^+\pi^-\pi^0)$ \cite{BABAR,CMD2}, $e^+e^- \to \omega\pi^+\pi^-\pi^0$ \cite{BABAR}, 
and for $e^+e^- \to 2(\pi^+\pi^-)\pi^0$ \cite{BABAR2} in the reaction cross section. 
Earlier measurements with lower statistics had already suggested the presence of such a dip, 
cf. the review \cite{Whalley}.

Phenomenological fits to the $\eebar$ data locate the structure at $1.91$ GeV \cite{FOCUS1} 
or at $1.86$--$1.88$ GeV \cite{BABAR} while the $\ppbar$ threshold is at $1.8765$ GeV.
Naturally, this very proximity of the $\bar NN$ threshold suggests that the $\bar NN$ channel 
should have something to do with the appearance of that structure in the multipion production cross
sections. A common speculation is that it could be a signal for a $\bar pp$ bound state 
\cite{FOCUS1,CMD} or a subtreshold $\bar pp$ resonance \cite{CMD2}. 
Such a conjecture seems to be also in line with experimental findings
in a related reaction, namely $\eebar \to \ppbar$, where a near-threshold enhancement 
seen in the cross section \cite{Aubert,Lees} is likewise associated with
a possible $\bar pp$ bound state, cf. also Ref.~\cite{Antonelli:1996}. 
For a discussion of other structures seen in 
$\eebar \to \ppbar$, see e.g. Ref.~\cite{Lorenz:2015pba}.

Given the complexity of the reaction mechanism one cannot expect a microscopic calculation
of those multipion production cross sections soon. Indeed, so far there is not even a
calculation that quantifies the impact of the opening of the $\bar NN$ channel on 
those reactions.
The idea that $e^+e^- \to 6\pi$ and $\eebar \to \ppbar$ could be closely related is 
picked up in Ref.~\cite{Russian} for interpreting the drop/dip. 
Aspects related to the question of an $\bar NN$ bound state are discussed in 
Ref.~\cite{Rosner} where it is emphasized that ordinary threshold effects like
cusps could also explain the structures seen in the multipion channels. 

In the present paper we investigate the significance of the $\bar NN$ channel for the 
$e^+e^- \to 5\pi$ and $e^+e^- \to 6\pi$ reactions. Specifically, we aim at a reliable
estimation of the influence of the $\bar NN$ channel on those multipion production cross 
sections around the $\bar NN$ threshold. The calculation builds on earlier works
published in Refs.~\cite{NNbar,NPA}. 
This concerns (i) an $\NNbar$ potential constructed in the framework of
chiral effective field theory (EFT), that reproduces the amplitudes determined
in a partial-wave analysis  of $\ppbar$ scattering data \cite{Timmermans}  
from the $\NNbar$ threshold up to laboratory energies of $T_{lab} \approx 200-250$ MeV \cite{NNbar} 
and (ii) an analysis of the reaction $\eebar \to \ppbar$ (and $\ppbar \to \eebar$, respectively) 
where the effect of the interaction in the $\NNbar$ system was taken into account 
rigorously \cite{NPA} and where the experimentally observed near-threshold enhancement in 
the cross section and the associated steep rise of the electromagnetic form factors in the 
time-like region is explained solely in terms of the $\ppbar$ interaction. 
Note that the employed $\NNbar$ interaction is also able to describe the large near-threshold 
enhancement observed in the reaction $J/\psi \to \gamma\ppbar$ \cite{Jpsi}. 

\section{Formalism}
The estimation of the influence of the $\bar NN$ channel is done in the same
framework as the studies mentioned above \cite{NNbar,NPA} and consistently with them. 
It amounts to solving the following formal set of coupled equations: 
\begin{eqnarray}
\nonumber
T_{\bar NN\to \bar NN} &=&  V_{\bar NN\to \bar NN} + V_{\bar NN\to \bar NN} G_0 T_{\bar NN\to \bar NN}, \\
\nonumber
T_{e^+e^- \to \bar NN} &=&  V_{e^+e^- \to \bar NN} + V_{e^+e^- \to \bar NN} G_0 T_{\bar NN\to \bar NN}, \\
\label{eq:LS0}
T_{\bar NN\to  \nu   } &=&  V_{\bar NN\to  \nu     } + T_{\bar NN\to \bar NN} G_0 V_{\bar NN\to  \nu   }, \\
\nonumber
& & \\ \nonumber 
T_{e^+e^- \to  \nu   } &=&  A_{e^+e^- \to  \nu} + V_{e^+e^- \to \bar NN} G_0 T_{\bar NN\to  \nu   }  \\
                       &=&  A_{e^+e^- \to  \nu} + T_{e^+e^- \to \bar NN} G_0 V_{\bar NN\to  \nu   },   
\label{eq:LS}
\end{eqnarray}
with $G_0$ the free $\NNbar$ propagator and $\nu =3(\pi^+\pi^-)$, etc. 
Here the first one is the Lippmann-Schwinger equation from which the $\NNbar$ scattering amplitude
is obtained, see Ref.~\cite{NNbar} for details. 
The second equation provides the $e^+e^- \to \bar NN$ transition amplitude, which was calculated in 
distorted-wave Born approximation in Ref.~\cite{NPA}. The third equation defines the amplitude 
for $\NNbar$ annihilation into the (various) $5 \pi$ and $6 \pi$ channels. These amplitudes will be 
established in the present work. Luckily there is experimental information for all considered multipion 
channels, i.e. for $\ppbar \to 3(\pi^+\pi^-)$, $\ppbar \to 2(\pi^+\pi^-\pi^0)$, $\ppbar \to 2(\pi^+\pi^-)\pi^0$, 
and $\ppbar \to \omega \pi^+\pi^-\pi^0$ \cite{Klempt,McCrady,Sai}, so that the corresponding 
transition potentials $V_{\bar NN\to\nu }$ can be constrained by a fit to data. 
Once $T_{e^+e^- \to \bar NN}$ ($V_{e^+e^- \to \bar NN}$) and $T_{e^+e^- \to \nu }$ ($V_{e^+e^- \to \nu }$)
are fixed the contributions to the $e^+e^- \to \nu$ reactions that proceed via an
intermediate $\NNbar$ state, cf. the second terms on the right hand side of Eq.~(\ref{eq:LS}), 
are likewise fixed. Note that the two lines in Eq.~(\ref{eq:LS}) are equivalent. 
The only unknown quantity in the equations above is $A_{e^+e^- \to\nu }$.
It stands for all other contributions to $e^+e^- \to \nu$, i.e. practically speaking it represents  
the background to the loop contribution due to two-step $e^+e^- \to \NNbar \to \nu$ transition.

Of course, it is impossible to take into account the full complexity of the 
$\eebar \to  5 \pi $, $\eebar \to  6 \pi $ and $\NNbar \to  5 \pi $, $\NNbar \to  6 \pi $ reactions. 
Thus, in the following we want to describe the simplifications and approximations made in our study. 
First of all this concerns the reaction dynamics. Following the strategy in Ref.~\cite{NNbar}, the 
elementary transition (annihilation) potential for $\NNbar \to\nu$ is parameterized by  
\begin{equation} 
V_{\NNbar \to \nu}(q)=\tilde C_\nu + C_\nu q^2,
\label{eq:VNN}
\end{equation}
i.e. by two contact terms analogous to those that arise up to next-to-next-to-leading order (NNLO) 
in the treatment of the $\NNbar$ or $NN$ interaction within chiral EFT. The quantity $q$ in 
Eq.~(\ref{eq:VNN}) is the center-of mass (c.m.) momentum in the $\NNbar$ system. Since the threshold 
for the production of $5$ or $6$ pions lies significantly below 
the one for $\NNbar$ the pions carry - on average - already fairly high momenta. Thus, 
the dependence of the annihilation potential on those momenta should be small for energies around 
the $\NNbar$ threshold and it is, therefore, neglected. 
The constants $\tilde C_\nu$ and $C_\nu$ are determined by a fit to the $\NNbar \to \nu$
cross section (and/or branching ratio) for each annihilation channel $\nu$. 

The term $A_{\eebar \to\nu}$ is likewise parameterized in the form~(\ref{eq:VNN}), but as a 
function of the $\eebar$ c.m. momentum. The arguments for this simplification are the same as 
above and they are valid again, of course, only for energies 
around the $\NNbar$ threshold. However, since in the $\eebar$ case this term does represent 
actually a background amplitude and not a transition potential we allow the corresponding 
constants to be complex numbers which are fixed by a fit to the $\eebar \to\nu$ cross sections. 
 
In our study of the electromagnetic form factors in the time-like region \cite{NPA}
we adopted the standard one-photon approximation. In this case there are only two partial 
waves that can contribute to the $\eebar \to\NNbar$ transition, namely the 
(tensor) coupled ${^3S_1} - {}^3D_1$ partial waves. We make the same assumption in the
present work. For $\NNbar \to 5 \pi, 6\pi$ there are no general limitations on the 
partial waves. However, since we restrict ourselves to energies close to the
$\NNbar$ threshold and we expect the annihilation operator to be of rather short range 
any $\NNbar$ partial waves besides the ${^3S_1}$ and the ${^1S_0}$ should play a minor role.
In the actual calculation we use only the ${^3S_1}$ partial wave. Thus, the corresponding
transition potential $V_{\NNbar \to\nu} $ might actually overestimate the true contribution 
of this partial wave and, therefore, the resulting amplitude for 
the two step process $\eebar \to \NNbar \to\nu$ has to be considered as an upper 
limit. Judging from available branching ratios for $\ppbar \to \omega\omega$ \cite{Klempt},
where near threshold only the ${^1S_0}$ can contribute, its contribution (to the $2(\pi^+\pi^-\pi^0)$ 
channel) could be in the order of 20\% of the one from the ${^3S_1}$ partial wave.

Note that there are selection rules for the $\NNbar \to 5 \pi, 6\pi$ transitions, because $G$-parity is preserved.
For $n$ pions the $G$-parity is defined by $G=(-1)^n$ while for $\bar NN$ it is given by $G=(-1)^{L+S+I}$, 
where $L$, $S$, $I$ denote the orbital angular momentum, and the total spin and isospin, respectively. 
Thus, the $G$-parity for the six pion final states (i.e. also for $\omega \pi^+\pi^-\pi^0$) is positive 
which confines the isospin for the $\bar NN$ pair to $I=1$ in the ${^3S_1} - {}^3D_1$ partial wave. 
Conversely, the five-pion decay mode can occur only from the $I=0$ ${^3S_1} - {}^3D_1$ $\NNbar$ partial wave.

The explicit form of Eq.~(\ref{eq:LS}) reads
\begin{eqnarray}\label{eq:ee6piN}
&& T_{\nu, e^+e^-}(Q,q_e;E)= A_{\nu,e^+e^-}(Q,q_e) + \sum_{\NNbar} \int_0^\infty \frac{dq q^2}{(2\pi)^3} \nl 
&&\qquad \times V_{\nu,\NNbar}(Q,q) \frac{1}{E-2E_q+i0^+}T_{\NNbar,\eebar}(q,q_e;E), \nl
\end{eqnarray}
written here in matrix notation. The sum refers to $\ppbar$ and $\nnbar$ intermediate states. 
The corresponding expression for $\ppbar \to \nu$ can be obtained by substituting $\eebar$ by $\ppbar$
in Eq.~(\ref{eq:ee6piN}), those for the other amplitudes in Eq.~(\ref{eq:LS0}) are given in Refs.~\cite{NNbar,NPA}. 
The quantity $Q$ stands here symbolically for the momenta in the $5 \pi$ and $6 \pi$ channels. 
But since we assumed that the transition potentials do not depend on the pion momenta, 
cf. Eq.~(\ref{eq:VNN}), $Q$ does not enter anywhere
into the actual calculation and we do not need to specify this quantity. All amplitudes (and the potentials) 
can be written and evaluated as functions of the c.m. momenta in the $\NNbar$ ($q_p$) and $\eebar$ ($q_e$)
systems and of the total energy $E=2 \sqrt{m_p^2 + q^2_p} = 2 \sqrt{m_e^2 + q^2_e} $.   
The quantity $E_q$ in Eq.~(\ref{eq:ee6piN}) is given by $E_q= \sqrt{m_p^2 + q^2}$. 

Since the amplitudes do not depend on $Q$ the integration over the multipion phase space can
be done separately when the cross sections are calculated. In practice, it amounts only to a
multiplicative factor and, moreover, to factors that are the same for $\eebar \to \nu$ and 
$\NNbar \to \nu$ at the same total energy $E$. We performed this phase space integration
numerically at the initial stage of the present work but it became clear that we can get
more or less equivalent results if we simulate that multipion phase space by 
effective two-body channels with a threshold that coincides with the ones of the
multipion systems. In effect the differences in the phase space can be simply absorbed 
into the constants in the transition potentials, see Eq.~(\ref{eq:VNN}) --
which anyway have to be fitted to the data. All results presented in this manuscript
are based on an effective two-body phase space. 

Of course, this simulation via effective two-body channels works only for energies around
the $\NNbar$ threshold. We cannot extend our calculation down to the threshold of the multipion
channels. However, one has to keep in mind that also the validity of our $\NNbar$ interaction
is limited to energies not too far away from the $\NNbar$ threshold. Thus, we have to
restrict our study to that small region around the threshold anyway. 

With the definitions of the $T$-matrices above, the cross section is obtained via 
\begin{equation}
\sigma_{\eebar\to\nu} (E)=\frac{3 E^2 \beta}{2^{10}\pi^3}\,|T_{\nu,\eebar}(E)|^2,
\end{equation}
and similarly for $\ppbar\to\nu$. The quantity $\beta$ denotes the phase space factor for 
an effective two-body system with equal masses $M$, $\beta=\sqrt{(E^2-4M^2)}/\sqrt{(E^2-4m_e^2)}$,
with $2M=6m_\pi$, $5m_\pi$, or $m_\omega3m_\pi$. For $\ppbar\to\nu$ the 
electron mass ($m_e$) has to be replaced by the one of the proton. 

\section{The $\NNbar \to 5\pi, 6\pi$ reactions}
First we need to fix the constants $\tilde C_\nu$ and $C_\nu$ in the $\NNbar \to \nu$
transition potentials. We do this by considering available branching ratios 
of $\ppbar$ annihilation at rest for the $3(\pi^+\pi^-)$, $2(\pi^+\pi^-\pi^0)$,
$\omega \pi^+\pi^-\pi^0 $, and $2(\pi^+\pi^-)\pi^0$ channels \cite{Klempt,McCrady}. 
For the annihilation into $3(\pi^+\pi^-)$ and $2(\pi^+\pi^-)\pi^0$ there are, in addition,
in flight measurements for energies not too far from the $\NNbar$ threshold
\cite{Sai}. Following Ref.~\cite{Juelich_NNbar} we evaluate the relative cross sections 
at low energy and compare those with the measured branching ratios. Specifically, we 
calculate the cross sections at $p_{\rm lab} = 106$ MeV/c ($T_{\rm lab} \approx 5$ MeV) 
because at this energy the total annihilation cross section is known from 
experiment \cite{BRrest} and it can be used to calibrate the cross sections for the 
individual $5 \pi$ and $6\pi$ channels based on the branching ratios. 

\begin{figure}[htbp]
\begin{center}
\includegraphics[height=110mm,clip]{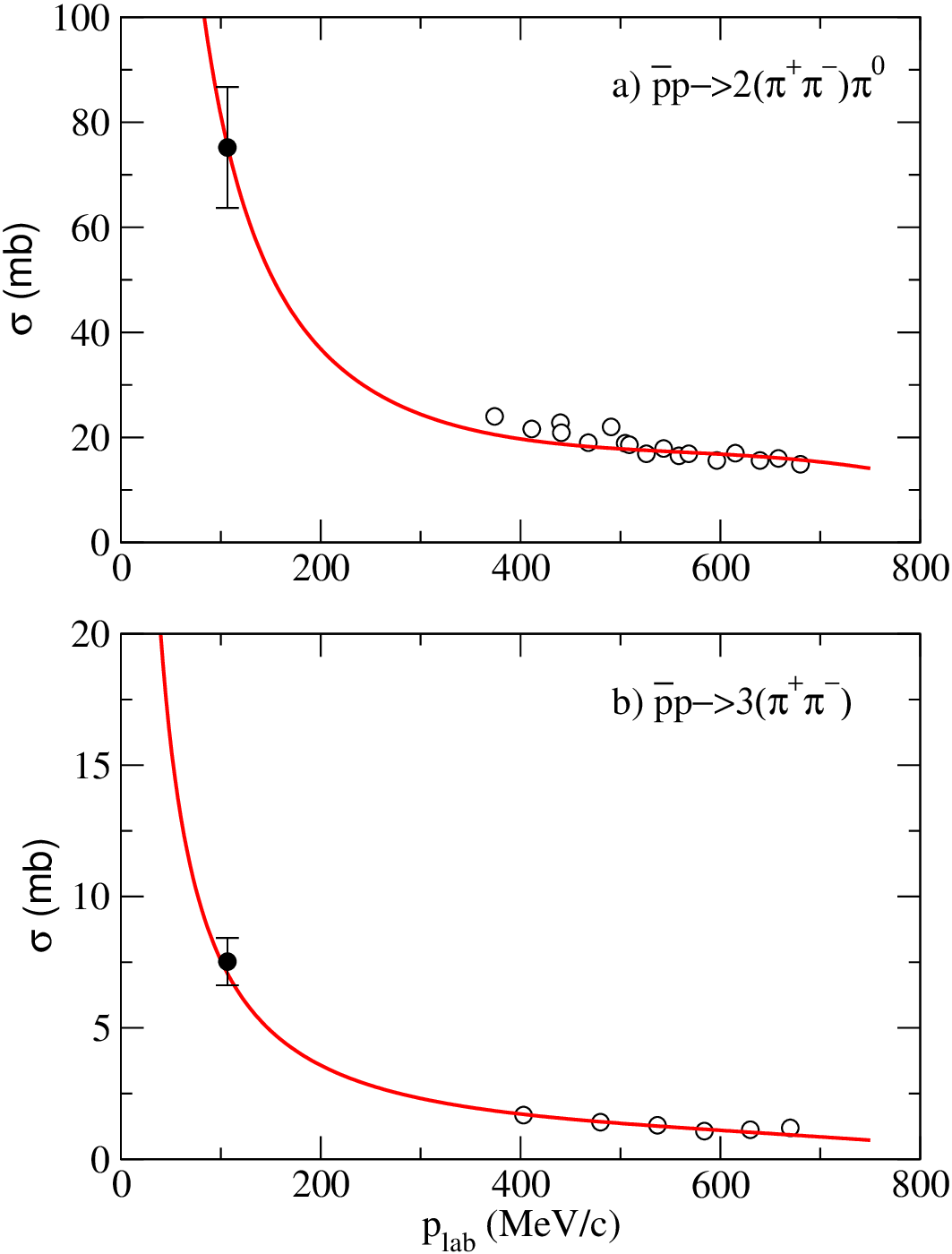}
\caption{(Color online) Cross section for (a) $\bar pp\to 2(\pi^+\pi^-)\pi^0$ 
and (b) $\ppbar\to 3(\pi^+\pi^-)$.
The solid  curves represent our result. Data are taken from Ref.~\cite{Sai} (open circles).  
The ``data'' points at $106$~MeV/c (filled circles) are deduced from 
information on the branching 
ratios of $\bar pp$ annihilation at rest, see text. 
}
\label{fig:ppbar6pi}
\end{center}
\end{figure}

Our results for $\ppbar\to 3(\pi^+\pi^-)$ and $\bar pp\to 2(\pi^+\pi^-)\pi^0$
are shown in Fig.~\ref{fig:ppbar6pi}. Note that the lowest ``data'' point is not 
from a measurement but deduced from the branching ratios \cite{Klempt} and 
the total annihilation cross section \cite{BRrest} as discussed in the preceding
paragraph. 

There are no in flight data for $\ppbar \to 2(\pi^+\pi^-\pi^0)$ and 
$\ppbar \to \omega \pi^+\pi^-\pi^0$. Here we fit to the central value of the branching 
ratios, 17.7\% \cite{Klempt} and 16.1\% \cite{McCrady}, respectively, and assume
that the energy dependence is the same as for the $3(\pi^+\pi^-)$ channel. The resulting
cross sections at $p_{\rm lab} = 106$ MeV/c are $63.2$ mb for the $2(\pi^+\pi^-\pi^0)$ 
case and $57.5$ mb for $\omega 3\pi$. 
Note that the uncertainty in the energy dependence is not too critical. 
Important is first and foremost the absolute value of those cross sections close to 
the $\NNbar$ threshold because that value is decisive for the magnitude of the
$\eebar \to \NNbar \to \nu$ two-step contribution and, in turn, for the relevance
of the $\NNbar$ intermediate state for the $\eebar \to \nu$ reaction. 

The results above are based on the NNLO EFT $\NNbar$ interaction with the cutoff 
combination $\{\Lambda, \tilde\Lambda\} = \{450,500\}$ MeV, cf. Ref.~\cite{NNbar}
for details. Exploratory calculations for the other cutoff combinations considered
in Ref.~\cite{NNbar} turned out to be very similar. Like for $\NNbar$
scattering itself, much of the cutoff dependence is absorbed by the contact terms
($\tilde C_\nu$ and $C_\nu$ in Eq.~(\ref{eq:VNN})) that are fitted to the data so that the
variation of the results for energies of, say, $\pm 50$ MeV around the $\NNbar$
threshold is rather small. For consistency the momentum dependent regulator
function as given in Eq.~(2.21) in Ref.~\cite{NNbar} is also attached to all 
momentum dependent quantities here, for example to the transition potential in
Eq.~(\ref{eq:VNN}). 

Because of the coupled nature of the $^3S_1$-$^3D_1$ $\NNbar$ partial wave, in principle, 
the $D$ wave should be also included in Eq.~(\ref{eq:ee6piN}) and, consequently, also in
Eq.~(\ref{eq:VNN}). Then there would be an additional contact term of the form 
$D_\nu q^2$ \cite{NNbar}, representing the $\NNbar \to \nu$ transition potential from 
the $\NNbar$ $^3D_1$ state, and a summation over the intermediate $\NNbar$ $^3D_1$ 
state arises, in addition to the integration over the intermediate 
momentum in Eq.~(\ref{eq:ee6piN}). 
We ignore these complications here because transitions starting from the 
$\NNbar$ $D$ wave are strongly suppressed for energies around the $\NNbar$ threshold 
and the contribution from the loop can be anyway effectively included in the contact
terms of the transition from the $\NNbar$ $S$-wave state. 

\section{Results for $\eebar \to 5\pi, 6\pi$}
Once the contact terms in $V_{\NNbar \to \nu}$ are fixed from a fit to the pertinent
data the corresponding part of the ${\eebar \to \nu}$ amplitude that comes from the
transition via an intermediate $\NNbar$ state is also completely fixed, cf. Eq.~(\ref{eq:LS}). 
We then add $A_{\eebar\to\nu}$. This term is assumed to be of the same functional 
form as Eq.~(\ref{eq:VNN}), however, it can no longer be 
identified with a transition potential (like for ${\NNbar \to \nu}$) but rather has to 
account for all other contributions to ${\eebar \to \nu}$, besides the one that includes
the intermediate $\NNbar$ state. Specifically, this term can have a relative phase as
compared to the contribution from the $\NNbar$ loop. Therefore, in this case the 
parameters can and should be complex. Since this background amplitude simulates a possibly 
very large set of transition
processes it should have a weak dependence on the total energy in the region of the $\NNbar$
threshold and this feature is implemented by the ansatz~(\ref{eq:VNN}) with $q$ being interpreted 
as the c.m. momentum in the ${\eebar}$ system. The two complex constants in the 
analogous Eq.~(\ref{eq:VNN}) for $A_{\eebar\to\nu}$
are adjusted in a fit to the cross sections of each of the four ${\eebar \to \nu}$ reactions
studied in the present investigation. For the fit we considered data in the range
$1750 \ \rm{MeV}  \le E \le 1950$~MeV, i.e. in a region that spans more or less symmetrically the $\NNbar$ threshold. 

\begin{figure}
\begin{center}
\includegraphics[height=60mm,clip]{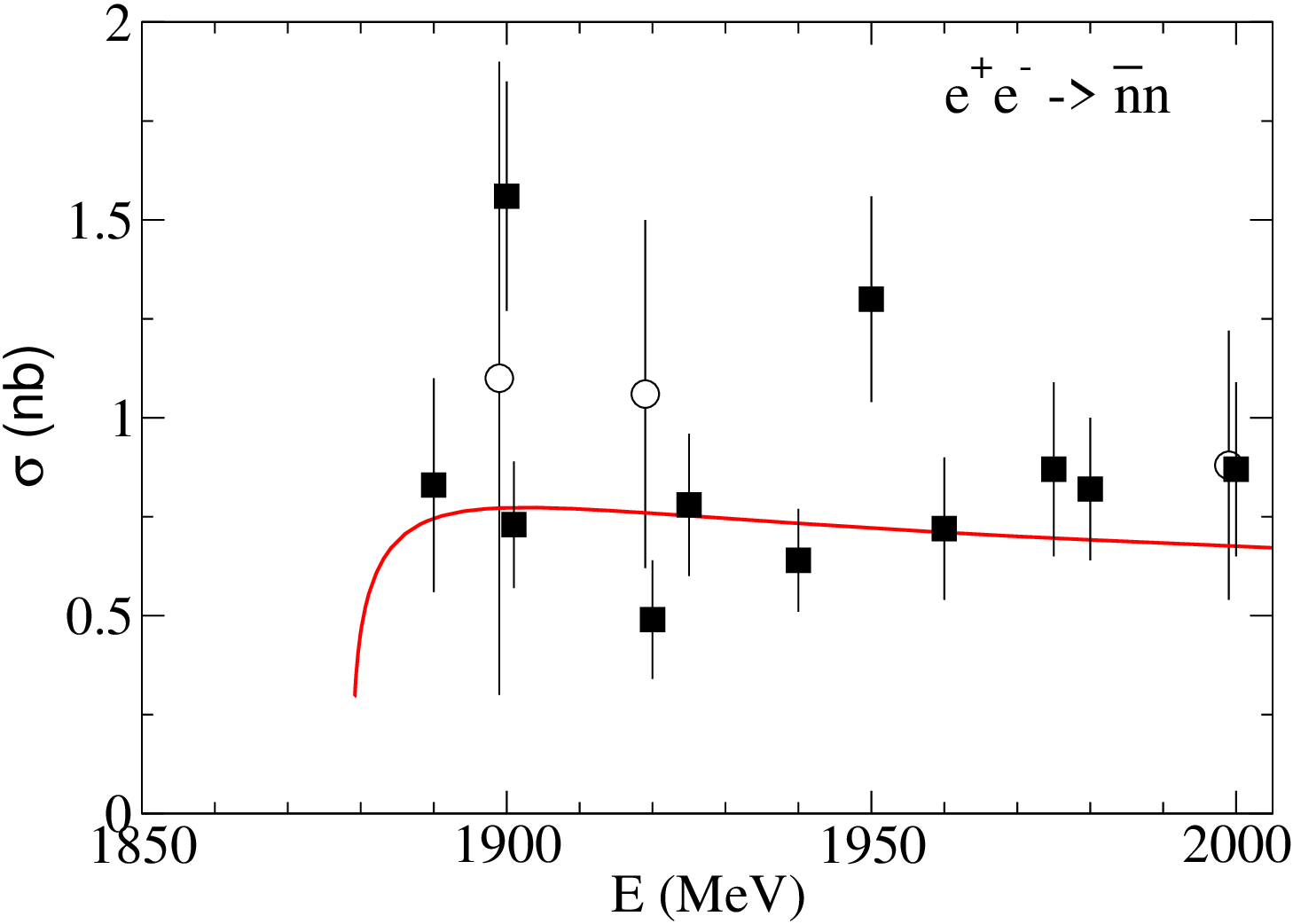}
\caption{(Color online) Cross section for $\eebar\to\nnbar$. The solid line represents our result.
Data are taken from Refs.~\cite{Antonelli} (circles) and \cite{Xsec_nnbar} (squares). 
}
\label{fig:nnbar}
\end{center}
\end{figure}

In principle, the ${\eebar \to \ppbar}$ amplitude that enters the loop contribution in 
Eq.~(\ref{eq:ee6piN}) can be taken straight from Ref.~\cite{NPA} where it was fixed 
in a fit to the ${\eebar \to \ppbar}$ cross section. The results in that work 
were obtained by using a $\ppbar$ amplitude which is the sum of the isospin $I=0$ and
$I=1$ amplitudes, i.e. $T_{\ppbar} = (T^{I=1}+T^{I=0})/2$. However, it was found 
that employing other combinations of $T^{1}$ and $T^{0}$ lead to very similar 
results and in all cases an excellent agreement with the energy dependence exhibited
by the data could be achieved. Thus, since isospin is not conserved in the reaction 
${\eebar \to \ppbar}$ the actual isospin content of the produced $\ppbar$ could not be fixed. 
The mentioned selection rules for $\NNbar \to n\,\pi$ imply that the $6\pi$ final state
can only be reached from an $I=1$ $^3S_1$ $\NNbar$ state while $5$ pions have to come
from the corresponding $I=0$ state. Thus, the magnitude of the $\NNbar$ loop contribution
to ${\eebar \to \nu}$ depends decisively on the isospin content of the intermediate
$\NNbar$ state. We did calculations for ${\eebar \to \nu}$ with the combination as used in Ref.~\cite{NPA} 
but it turned out that a slightly larger $I=1$ admixture, namely $T_{\bar pp}\approx 0.7\,T^1+0.3\,T^0$,
is preferable and leads to a somewhat better overall agreement with the experiments and, therefore,
we adopt this combination here.  
The cross section for ${\eebar \to \nnbar}$ is also known experimentally \cite{Antonelli,Xsec_nnbar},
though with somewhat less accuracy. It agrees with the one for ${\eebar \to \ppbar}$ within the error bars
\cite{Xsec_nnbar}. 
Therefore, we simply put $T_{\eebar \to \nnbar}=T_{\eebar \to \ppbar}$ in the sum in Eq.~(\ref{eq:ee6piN}),  
which is certainly justified as can be seen from the actual ${\eebar \to \nnbar}$ result presented in 
Fig.~\ref{fig:nnbar}. 

As discussed above, $D$-wave contributions were ignored in case of $\ppbar\to \nu$. However, for
$\eebar\to \nu$ around the $\NNbar$ threshold the momentum in the incoming system is no longer 
small and the $\eebar$ $^3D_1$ component cannot be neglected. However, it can be easily included
because the $\eebar \to \ppbar$ transition amplitudes from the $^3S_1$ and $^3D_1$ $\eebar$
states are proportional to each other, see Eq.~(6) of Ref.~\cite{NPA}. Thus, for including
the $D$-wave contribution we simply have to multiply the $S$-wave cross section by a 
factor $1.5$. 

\begin{figure*}[htbp]
\begin{center}
\includegraphics[height=120mm,clip]{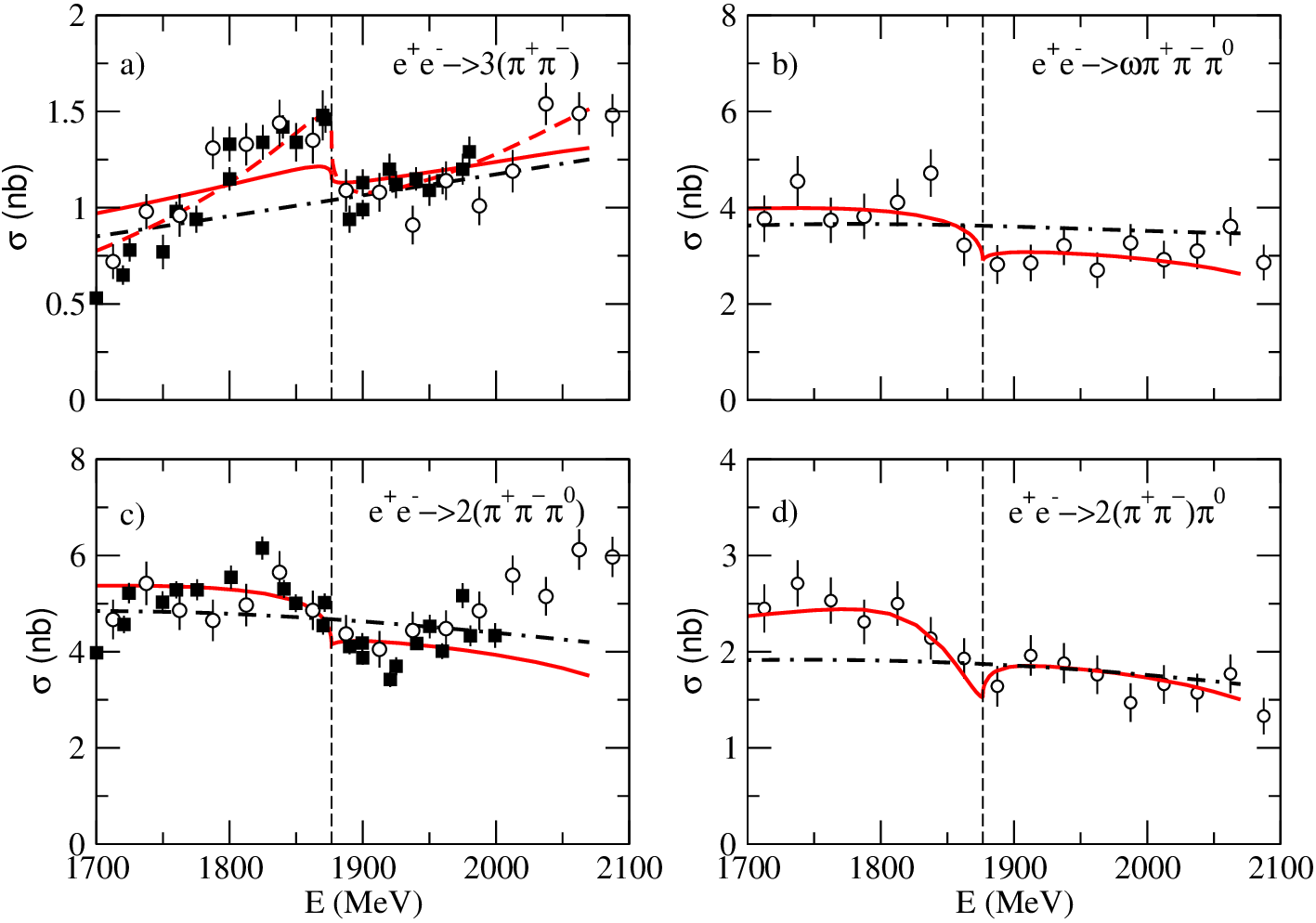}
\caption{(Color online) Cross section for (a) $e^+e^-\to 3(\pi^+\pi^-)$, 
(b) $\omega\pi^+\pi^-\pi^0$, 
(c) $2(\pi^+\pi^-\pi^0)$, and (d) $2(\pi^+\pi^-)\pi^0$.
The solid (red) curves represent our full result, including the $\NNbar$ intermediate state, 
while the dash-dotted (black) curves are based on the background term alone. 
The vertical lines indicate the $\bar pp$ threshold.
The dashed (red) curve in (a) corresponds to amplifying deliberately the $\bar NN$ loop 
contribution by a factor of four.
Data are taken from Refs.~\cite{BABAR,BABAR2} (circles) and \cite{CMD,CMD2} (squares). 
}
\label{fig:ee6pi}
\end{center}
\end{figure*}

Our results are shown in Fig.~\ref{fig:ee6pi}.
Obviously, in three of the four considered reactions the contribution from the two-step
process $\eebar\to \NNbar\to\nu$ is large enough to be of relevance and, moreover, together
with a suitably adjusted background a rather good description of the cross sections around
the $\NNbar$ threshold can be achieved (solid curves). The cross section due to the background 
alone is indicated by the dash-dotted curves. It is practically constant and does not exhibit
any structure. The contribution involving the intermediate $\NNbar$ state generates a distinct
structure at the $\NNbar$ threshold and is responsible for the fact that the full result is
indeed in line with the behaviour suggested by the measurements. Of course, 
in case of the $2(\pi^+\pi^-\pi^0)$ channel the data could hint at a minimum at an 
energy slightly above the threshold. However, there is also some variation between the
two experiments. Further measurements with improved statistics and also with a better  
momentum resolution would be quite helpful. This applies certainly also to the other channels.

No satisfactory result could be achieved for the reaction $e^+e^-\to 3(\pi^+\pi^-)$. Here the
amplitude due to the intermediate $\NNbar$ state would have to be roughly a factor four
larger in order to explain the data, see the dashed line. We emphasize that this
curve is shown only for illustrative purposes! At the moment we do not have any physical
arguments why that particular amplitude should be increased by a factor four. Indeed, we have
examined and explored various uncertainties that could be used to motivate an
amplification of the amplitude but without success. 
For example, assuming that the $\eebar\to \ppbar$ reaction is given 
by the isospin $1$ alone changes the result only marginally and the same is the case if 
we take into account that the $\eebar\to \nnbar$ cross section could by slightly larger than the 
one for $\eebar\to \ppbar$ as indicated by the data in Ref.~\cite{Antonelli}. 

\begin{figure}[htbp]
\begin{center}
\includegraphics[height=110mm,clip]{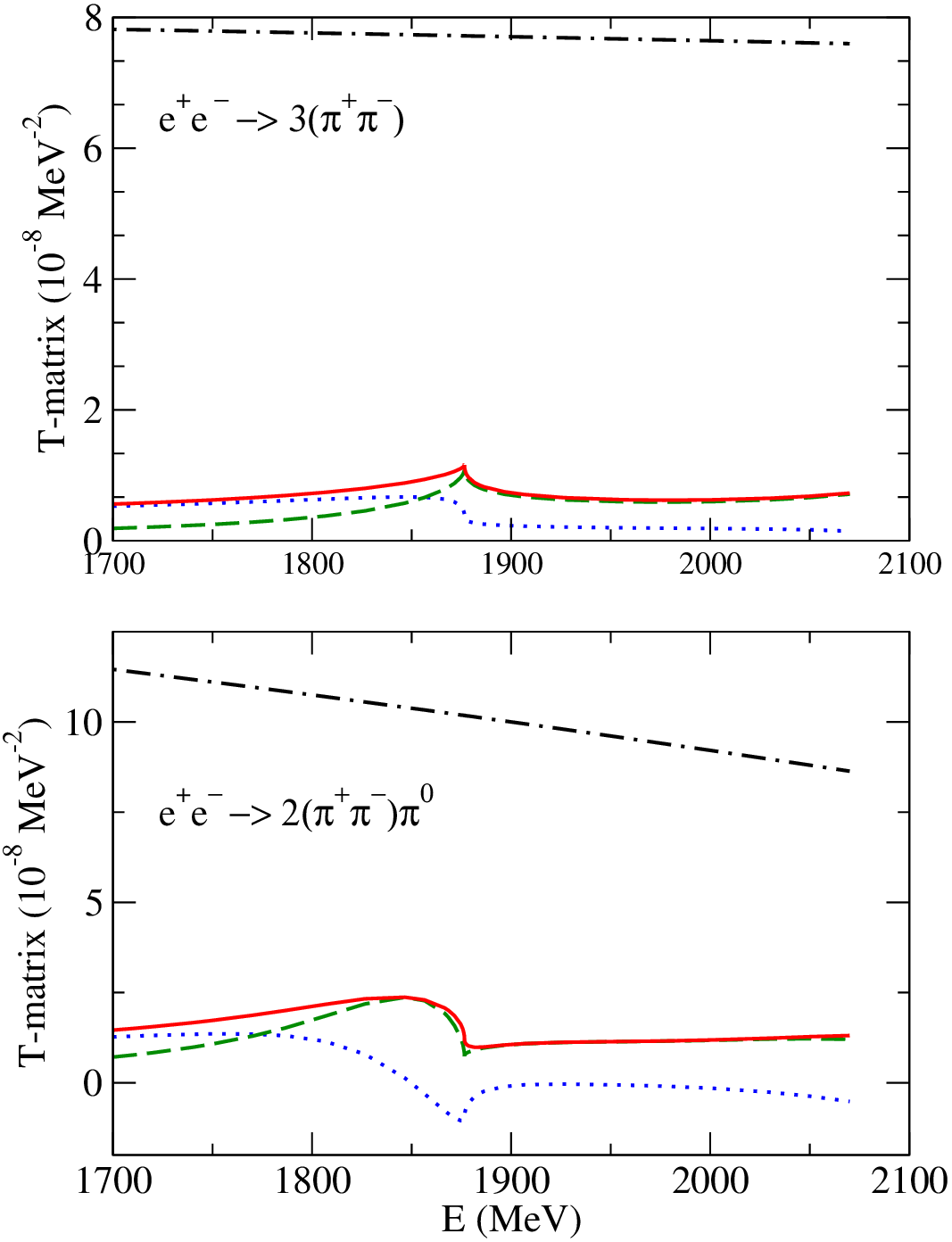}
\caption{(Color online) Real (dotted/blue curve) and imaginary part (dashed/green), and 
modulus (solid/red line) of the $\bar NN$ loop contribution, see Eq.~\eqref{eq:ee6piN}.
The modulus of the background term is shown by the dash-dotted (black) curves. 
}
\label{fig:ppbar6piA}
\end{center}
\end{figure}

Finally, let us come to the key question, namely are those structures seen in the 
experiment a signal for a $\NNbar$ bound state? As discussed in Ref.~\cite{NNbar}, 
we did not find any near-threshold poles for our EFT $\NNbar$ interaction in the
$^{3}S_1$--$^{3}D_1$ partial wave with $I=1$.  However, there is a pole in 
the $I=0$ case and this pole corresponds to a ``binding'' energy of 
$E_B= (+4.8-{\rm i}\,68.2.9)$ MeV for the NNLO interaction employed
in the present study \cite{NNbar}. 
The positive sign of the real part of $E_B$ indicates that the pole
we found is actually located above the $\bar NN$ threshold (in the energy plane). 
As discussed in Ref.~\cite{NNbar}, the pole moves below the threshold when we switch 
off the imaginary part of the potential and that is the reason why we refer to it as bound state.

There is a distinct difference in the $\eebar\to\nu$ amplitudes due to the $\NNbar$
loop contribution for the two isospin channels, see Fig.~\ref{fig:ppbar6piA}, and
the modulus exhibits indeed the features one expects in case of the absence/presence 
of a bound state, namely a genuine cusp or a rounded step and a maximum below the 
threshold. 
However, the structure in the cross section is strongly influenced and 
modified by the interference with the (complex) background amplitude as
testified by the results in Fig.~\ref{fig:ee6pi}. Thus, at this
stage we do not see a convincing evidence for the presence of an $\NNbar$
bound state in the data for $2(\pi^+\pi^-)\pi^0$ and for the opposite 
in case of $\omega\pi^+\pi^-\pi^0$, say. However, high accuracy data around
the $\NNbar$ threshold together with a better theoretical understanding of the background
could certainly change the perspective for more reliable conclusions in the future.

In any case, our results corroborate that one should see an effect of the opening
of the $\NNbar$ channel in the cross sections of the considered $\eebar\to\nu$
reactions. Thus, the observation of a dip or a cusp-like structure in that
energy region is not really something unusual or exotic. 
As argued above, our calculation provides a fairly reliable estimate for the 
amplitude that results from two-step processes with an intermediate $\NNbar$ state. 
Though all the reactions considered in the present study are obviously dominated by 
processes that are not related to the $\NNbar$ interaction, cf.
Fig.~\ref{fig:ppbar6piA}, the amplitude due to the
coupling to the $\NNbar$ system is large enough so that it can produce sizeable 
interference effects. In three of the four reactions investigated those 
interference effects are indeed sufficient to explain the behavior of the measured
cross sections in the region around the $\NNbar$ threshold. 

Should one expect similar structures also in other annihilation channels such as
$\eebar \to \pi^+\pi^-$, $\eebar \to 2(\pi^+\pi^-)$, etc.? An educated guess can be 
made based on the relative magnitude of the branching ratios for the pertinent $\NNbar$ 
annihilation channels compared to annihilation cross sections from the $\eebar$ state. 
Judging from the branching ratios summarized in Ref.~\cite{Klempt} 
and the compilation of $\eebar$ induced cross sections in Refs.~\cite{Druzhinin,Bevan}
the most promising case is certainly the $\pi^+\pi^-\pi^0 $ channel, see Table~\ref{tab:N1}. 
In fact, the available data \cite{Aubert:2004,Logashenko} could hint at an anomaly around the 
$\NNbar$ threshold. In case of the 4$\pi$ channels there is a kink at the $\NNbar$ threshold,
see, e.g. Ref.~\cite{Logashenko}. On the other hand, for the $\pi^+\pi^-$ case the
branching ratio is very small, see Table~\ref{tab:N1}, so that we do not expect any
noticeable effects there. Indeed, the data for $\eebar \to \pi^+\pi^-$ \cite{Aubert:2009} 
support this conjecture.

\begin{table}
\caption{Branching ratios for $\ppbar$ annihilation at rest \cite{Klempt}
and ${\eebar}$ annihilation cross sections around the $\NNbar$ threshold \cite{Druzhinin,Bevan}. 
}
\renewcommand{\arraystretch}{1.3}
\label{tab:N1}
\vspace{0.2cm}
\centering
\begin{tabular}{|c||c|c|}
\hline
$\nu$  & \ BR for $\ppbar \to \nu$ [\%] \ &  \ $\sigma_{\eebar \to \nu}$ [nb] \ \\
\hline
$\pi^+\pi^- $        & 0.314$\pm$0.012   &  $\approx$\,1 \\
$\pi^+\pi^-\pi^0 $    &  6.7$\pm$1.0    &  $\approx$\,1 \\
$2(\pi^+\pi^-)    $  &  5.6$\pm$0.9     &  $\approx$\,6 \\
$\pi^+\pi^-2\pi^0  $ & 12.2$\pm$1.8     &  $\approx$\,9 \\
$2(\pi^+\pi^-)\pi^0$ & 21.0$\pm$3.2     &  $\approx$\,2 \\
$\,2(\pi^+\pi^-\pi^0)\,$ & 17.7$\pm$2.7     &  $\approx$\,4 \\
$3(\pi^+\pi^-)     $ &  2.1$\pm$0.25    &  $\approx$\,1 \\
\hline
\end{tabular}
\renewcommand{\arraystretch}{1.0}
\end{table}

\section{Summary}
We analyzed the origin of the structure observed in the reactions $e^+e^-\to 3(\pi^+\pi^-)$, $2(\pi^+\pi^-\pi^0)$, 
$\omega\pi^+\pi^-\pi^0$, and $e^+e^-\to 2(\pi^+\pi^-)\pi^0$ around the 
$\bar pp$ threshold in recent BaBar and CMD measurements. Specifically, 
we evaluated the contribution of the two-step process $e^+e^-\to \bar NN \to$ multipions to the 
total reaction amplitude. The amplitude for $e^+e^-\to \bar NN$ was constrained from near-threshold 
data on the $e^+e^-\to \bar pp$ cross section and the one for $\bar NN \to$ multipions was fixed from 
available experimental information, for all those $5 \pi$ and $6\pi$ states.
The resulting amplitude turned out to be large enough to play a role for the considered $e^+e^-$ annihilation
channels and, in three of the four reactions, even allowed us to reproduce the data quantitatively near the
$\bar NN$ threshold once the interference with a background amplitude was taken into account. The
latter simulates other transition processes that do not involve an $\bar NN$ intermediate state. 
In case of the reaction $e^+e^-\to 3(\pi^+\pi^-)$ there is also a visible effect from the $\bar NN$ 
channel, however, overall the magnitude of the pertinent amplitude is too small. 

In our study the structures seen in the experiments emerge as a threshold effect due to the opening 
of the $\bar NN$ channel. The question of a $\bar NN$ bound state is discussed, however, no firm
conclusion can be made. But it is certainly safe to say that the near-threshold behavior of 
the $e^+e^-\to \bar pp$ cross section and the structures seen in $e^+e^-\to 3(\pi^+\pi^-)$, etc. 
have the same origin. 

\section*{Acknowledgements}
We would like to thank Eberhard Klempt for help and comments regarding the branching
ratio measurements in $\ppbar$ annihilation and Evgeni P. Solodov for sending
us the BaBar data for $\eebar \to \omega\pi^+\pi^-\pi^0$. 
This work is supported in part by the DFG and the NSFC through
funds provided to the Sino-German CRC 110 ``Symmetries and
the Emergence of Structure in QCD''.


\end{document}